\documentstyle[pra,twocolumn,aps,epsfig]{revtex}
\begin{document}
\draft
\title{Atom holography }
\author{Oliver Zobay$^{1,2}$, Elena V. Goldstein$^{1}$ and Pierre Meystre$^1$ }
\address{$^1$Optical Sciences Center, University of Arizona, Tucson,
Arizona 85721\\ $^2$ Sussex Centre for Optical and Atomic Physics,
University of Sussex, Falmer, Brighton BN1 9QH, UK
\\ \medskip}\author{\small\parbox{14.2cm}{\small\hspace*{3mm}
We study the conditions under which atomic condensates can be used as
a recording media and then suggest a reading scheme which allows to
reconstruct an object with atomic reading beam. We show that good
recording can be achieved for flat condensate profiles and for
negative detunings between atomic Bohr frequency and optical field
frequency. The resolution of recording dramatically
depends on the relation between the healing length of the condensate and
the spatial frequency contents of the optical fields involved.
\\[3pt]PACS numbers: 03.75.-b 03.75.Fi 05.30.Jp 42.65.Hw}}
\maketitle

\section{Introduction}

In recent years, atom optics has been rapidly emerging as a new and exciting
subfield of atomic physics. The objective of atom optics is
to manipulate atomic beams in a way similar to conventional optics by
exploiting the wave properties of the atoms. Supported by advances in
laser technology and microstructure fabrication a number of significant
accomplishments have been realized in the laboratory with the demonstration
of, e.g., mirrors, lenses, and diffraction gratings for atomic beams
\cite{AdaSigMly94}.
A natural way of extending these studies consists in exploring the
possibilities of holographic imaging with atoms, the conventional optics
analogue of which has been well-known for several decades
\cite{Gab48,Gab49}.

Optical holography can be described as the three-dimensional
reconstruction of the optical image of an arbitrarily shaped object.
Typically, this is done in a two-step process where first the
information about the object is stored in a hologram. This hologram is
created by recording, e.g., with the help of a photographic film, the
interference pattern between scattered light originating from the illuminated
object and a (plane-wave) reference beam. The second step is the
reconstruction, which is performed by shining a reading
beam similar to the reference beam onto the hologram. The diffraction
of the reading beam from the recorded pattern yields a virtual
as well as a real optical image of the original object.

Drawing on this concept, the characteristic property of atomic
holography is that at least the final reading step is performed
with an atomic beam. In this way, an atom-optical image of the object is
created which in certain situations can be thought of as some sort
of material replica of the original. There are several reasons why the
realization of atom holography is of interest: From a basic point of view,
it significantly extends the already well-established range of analogies
between light and matter waves. But more importantly perhaps, it may also
have useful practical applications from atom lithography to the
manufacturing of microstructures, or quantum microfabrication.

One of the prerequisites for an actual implementation of atomic
holography is the availability of a reading beam of sufficient
monochromaticity and coherence. Given the rapid advances in atom optics
and especially in the realization of atom lasers, this requirement can be
expected to be met in the near future. Another important question concerns the
potentially detrimental influence of gravitional effects. One of the greatest
challenges, however, is the manufacturing of the actual hologram where
the information to be reconstructed is stored. Several schemes can be
considered. One possibility is to diffract the atoms from a mechanical mask.
The first successful realizations of such an approach have recently been
reported in Ref.\ \cite{MorYasKis96}. In these experiments the hologram was
manufactured as a binary mask written onto a thin silicon nitride
membrane. Such a hologram has the advantage of being
permanent, however, as the mask only allows for complete or
vanishing (binary) transmission of the beam at a given point one loses a
significant amount of information about the optical image. Another
interesting proposal was recently made in Ref.\ \cite{Sor97}. In
this setup the atomic beam is diffracted from the inhomogeneous light
field created by the superposition of object and reference beam. These
beams thus directly form the hologram.

The purpose of the present paper is to investigate the perspectives of
an alternative approach, namely the manufacturing of the hologram directly
within a Bose-Einstein condensate (BEC). Atomic Bose condensates have recently
been realized experimentally \cite{AndEnsMat95,DavMewJofAndKet95} and are
now available almost routinely in several laboratories. The motivation for
the present study is twofold. First, the possibility of such an ``all-atomic''
scheme is interesting in itself and deserves further examination.
Furthermore, it illustrates the wide potential applicability of condensates
in atoms optics as a tool to influence the trajectories of atoms from
external sources.

Our approach is based on two main ideas. The holographic information
is encoded into the condensate in the form of density modulations by using
writing and reference laser beams that form an optical potential for the
condensate atoms. As we show later on, the density modulations follow
closely the optical beam interference pattern if the condensate is of
sufficiently high density, i.e., if a Thomas-Fermi description is applicable
\cite{DalGioPit98}. All-atomic reading is then accomplished in a way
reminiscent of the Raman-Nath regime of diffraction between an atomic beam
and a light field.  \cite{AdaSigMly94} Specifically, the reading beam atoms,
that have a suitably chosen velocity, interact with the condensate atoms
via $s$-wave scattering and acquire a spatially dependent phase shift
reflecting the density modulations of the condensate. In the further
spatial propagation of the atoms, this phase shift gives rise to the formation
of the atom-optical image.

The proposed method is hence fundamentally different from a recent suggestion
to arbitrarily shape the center-of-mass wave function of an atom
(``wave-front engineering''). \ \cite{OlsDekHer98} Instead of pursuing an
holographic approach, this latter method makes use of a sequence of suitably
shaped laser pulses to obtain the desired wave front. In fact, our proposal
is more closely related to Ref. \cite{ChiFor98}, which also suggests using
of Bose-Einstein condensates to control particle deflection. Finally, we note
that the present work is also related to the discussion of
the analogy between matter-wave mixing phonomena in ultracold atomic samples
and conventional nonlinear optics, including in particular matter-wave
phase conjugation \cite{GolPlaMey95,GolPlaMey96,GolMey99} and four-wave mixing
\cite{LawPuBig98,TriBanJul98,Phi99}.

The paper is organized as follows. After briefly
recollecting the principles of optical holography Sec.\ II gives a
general discussion of our approach to atomic holography introduced
above. As an illustration in Sec.\ III the atom-optical imaging of a simple
object is worked out in detail. Summary and conclusions are given in
Sec.\ IV.

\section{Holographic imaging with atomic beams}

\subsection{Principles of optical holography}

The principles of optical holography in their most basic form are shown
in Fig.\ 1a; detailed expositions can be found, e.g., in
Ref.\ \cite{VelRey67}. An object is illuminated with a laser
wave and the resulting field $E_o({\bf r})$ is brought to interference
with the reference beam $E_r({\bf r})$. The ensuing superposition field
is recorded on a suitable medium, e.g., a photographic plate, in such
a way that the optical transmission of the medium becomes proportional
to the total field intensity
\begin{equation}
I=|E_r+E_o|^2=|E_o|^2+ |E_r|^2+E_r^\star E_o +E_rE_o^\star.
\label{inten}
\end{equation}
The wave front of a reading beam $E_{rd}({\bf r})$ impinging upon this
hologram is thus proportional to $I({\bf r})E_{rd}({\bf r})$
after it has penetrated the medium. In this expression,
the terms of interest are
$E_oE_r^\star E_{rd}+E_o^\star E_rE_{rd}.$
They contain the original object wavefront and its conjugate, and can be
used to construct a virtual and a real image of the object.
In optics several techniques have been developed which allow to
separately view each of these terms, such as side-band Fresnel holography,
Fraunhofer holography, Fourier transform holography, etc. \cite{VelRey67}.
For the present discussion of atom-optical holography we make use of
\centerline{\psfig{figure=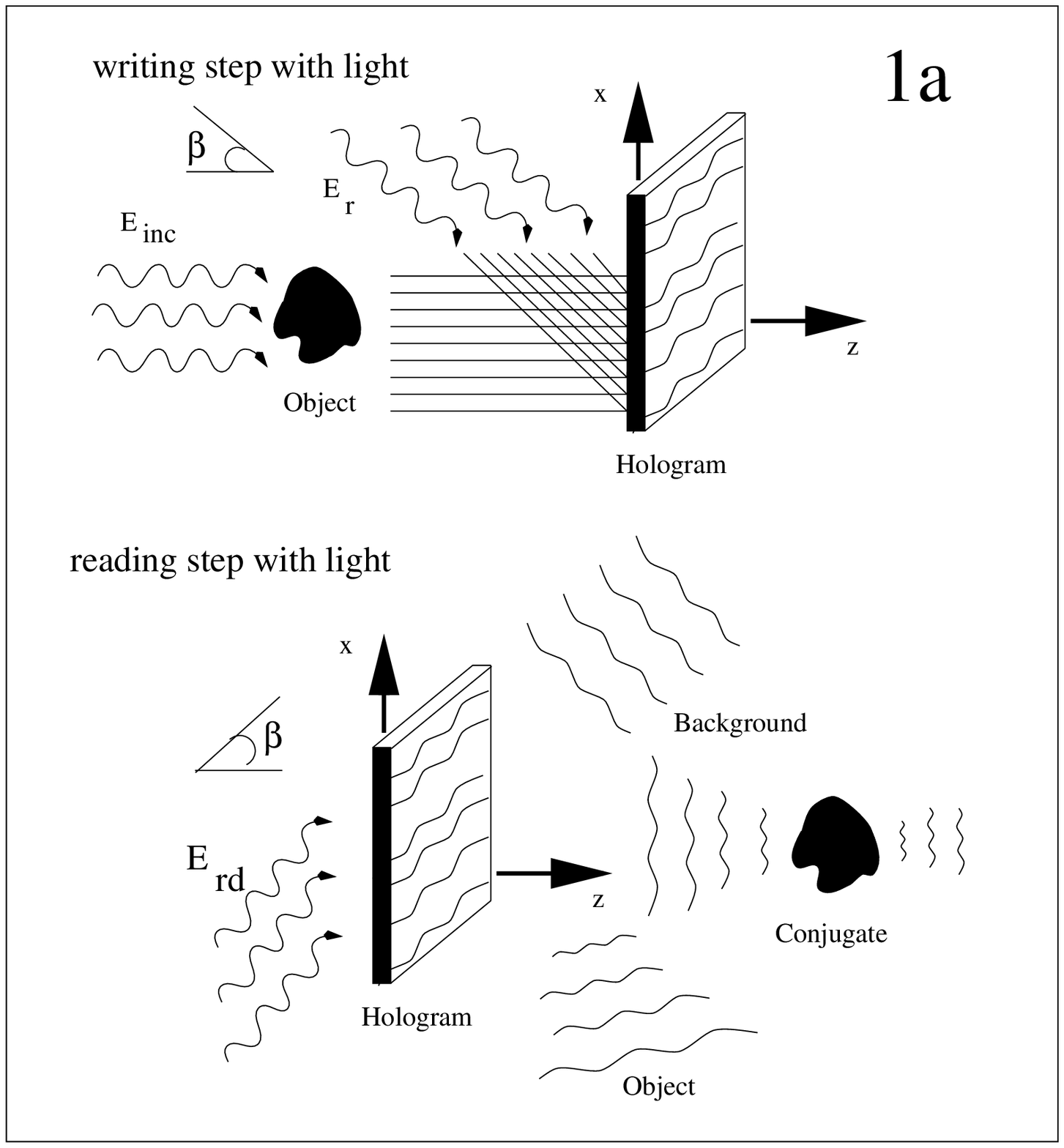,width=8.6cm,clip=}}
\centerline{\psfig{figure=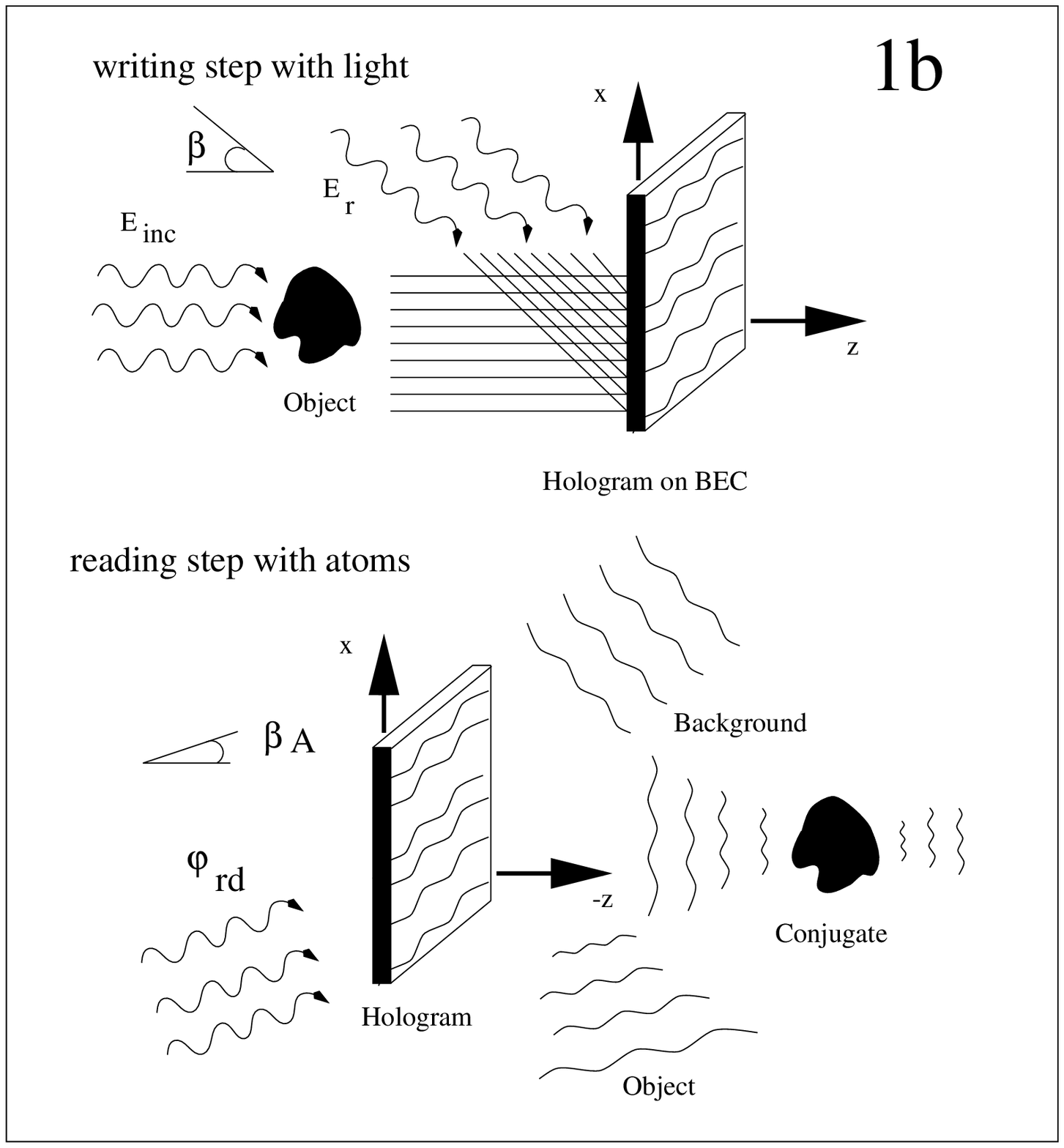,width=8.6cm,clip=}}
\begin{figure}
\caption{Set-ups for Fresnel side-mode
holography: (a) all-optical realization; (b) optical/matter-wave realization.
(Note the $-z$-axis in the last sketch.)}
\label{fig1}
\end{figure}
ideas from side-band Fresnel holography, which does not rely on lenses and
thus allows for a simple extension to matter waves.

\subsection{Atomic Bose-Einstein condensates as recording media}

The idea of storing information onto atomic condensates is based on the
observation that the density distribution of a condensate in the Thomas-Fermi
limit closely reflects the behavior of the confining potential. This yields
the possibility of accurate external control.

The Gross-Pitaevskii equation which governs the evolution of the
macroscopic wave function $\Phi({\bf r},t)$ describing the state of
an atomic condensate with $N$ atoms is given by \cite{DalGioPit98}
\begin{equation}
i\hbar \dot\Phi = \frac{{\bf p}^2}{2M}\Phi+V({\bf r})\Phi+
g|\Phi|^2\Phi,
\label{GP}
\end{equation}
where ${\bf p}$ denotes the atomic center-of-mass
momentum,  $M$ the atomic mass, and $V({\bf r})$ the external potential.
The strength of atomic two-body interactions is determined by
$g=4\pi\hbar^2 a/M$ with $a$ being the $s$-wave scattering length.
The normalization condition for the condensate wave function reads
\begin{equation}
\int d^3{\bf r}\,|\Phi({\bf r})|^2=N.
\label{norm}
\end{equation}

The steady state of a condensate can thus be described with a
time-independent wave function $\phi({\bf r})$ which is defined by
$$\Phi({\bf r},t)=e^{-i\mu t/\hbar}\phi({\bf r}),$$
where $\mu$ is the chemical potential.
In the Thomas-Fermi limit, where the effect of kinetic energy is much weaker
than the mean-field potential, the contribution of the term  ${\bf p}^2/2M$
can be neglected and the condensate density becomes
\begin{equation}
|\phi({\bf r})|^2=[\mu-V({\bf r})]/g,
\label{TFGS}
\end{equation} where $\mu$ is determined by the
normalization condition Eq.\ (\ref{norm}). From this expression we see
immediately that the form of the external potential is replicated
in the density profile of the atomic condensate.

Consider then replacing the photographic plate in Fig.1b by a pancake-shaped
BEC as the recording medium. In case the writing of the information into
the condensate is achieved by optical beams, which are assumed to be far 
detuned from atomic resonance, they create an optical potential proportional 
to $I/\delta$ where $I$ is given by Eq.\ (\ref{inten}) and 
$\delta=\omega-\omega_L$ is the detuning between the atomic resonance
$\omega$ and the laser frequency $\omega_L$. 
The total potential $V({\bf r})$ acting on the condensate
is then the sum of the trap potential, taken to be slowly varying,
and this optical potential. From Eq.\ (\ref{TFGS}), it then follows in
full analogy with optical holography that all terms in Eq.\ (\ref{inten})
are stored in the density distribution of the condensate ground
state. However, this atomic-condensate recording is in some ways more akin
to ``real-time'' holography, since the optical fields should be continuously
present in order to maintain the density modulations in the condensate.
\footnote{Note that the use of optical fields is not essential to the present
discussion: other interactions susceptible of imposing a spatially dependent
potential $V({\bf r})$ on the condensate can also be considered.}

\subsection{Reading from atomic condensates}

As already mentioned in the introduction, we consider an all-atomic reading
scheme, which has the fundamental advantage of allowing one to reconstruct
a {\em material} ``replica'' of the stored object. Specifically, the
reading beam is a monoenergetic atomic beam of velocity ${\bf v}_{rd}$
impinging at some angle onto the condensate. We assume that the
internal state of these incoming atoms is such that they are only weakly
perturbed by the writing and trap potentials, so that their dominant
interaction is scattering by the atoms in the condensate. It is important at
this point to emphasize that the atoms in the reading beam {\em need not}
be of the same species as the condensate atoms. In principle they
could be of just about any element or even molecule.

We consider specifically reading beam velocities such that the interaction
between the incoming atoms and the condensate can be described in terms
of $s$-wave scattering. This condition is fulfilled provided that \cite{LifPit80}
\begin{equation}
a_{rc}m_{rc}v_{rd}/\hbar \ll 1,
\label{scat_cond}
\end{equation}
where $a_{rc}$ denotes the $s$-wave scattering length for collisions
between reading and condensate atoms and $m_{rc}$ is their relative mass.
For simplicity, we further assume that the density of the reading beam is
low enough that collisions between atoms in that beam can be neglected.
Under these conditions the time evolution of the reading atoms' wave function
$\varphi({\bf r},t)$ in the mean field of the condensate is determined by
the equation
\begin{equation}
i\hbar \dot{\varphi}({\bf r},t)=\left[
\frac{{\bf p}^2}{2M_{rd}}+g_{rd}|\phi({\bf r})|^2\right]
\varphi({\bf r},t)
\label{Schr}
\end{equation}
where $M_{rd}$ is the mass of the atoms in the reading beam,
$g_{rd}=2\pi \hbar^2 a_{rc}/m_{rc}$. This equation
assumes that to a good degree of approximation, the condensate stays in
its ground state during the whole reading process.

Over the course of time, the condensate gradually loses atoms due to scattering
by the incoming atoms and other processes, but it is assumed that its
density distribution remains given by Eq.\ (\ref{TFGS}) with $\mu$
slowly varying due to the change in the number of atoms $N$, so that
$|\phi({\bf r})|^2$ has to be changed adiabatically in Eq.\ (\ref{Schr}).
Under these circumstances the shape of the holographic image will gradually change
and eventually distort when $|\phi({\bf r})|^2$ deviates too
much from the Thomas-Fermi expression. However, the time scale for this
process, the lifetime of the condensate, can be long in comparison to the
time necessary to form the image, the flight time of the reading atoms.

The reading and reconstruction of the condensate information into a
material ``replica'' is easily achieved if the condensate is sufficiently
thin that its density distribution can be regarded as effectively
two-dimensional, and its interaction with the reading atoms is short enough that
the Raman-Nath (or thin hologram) approximation can be invoked. The condensate
then acts as a phase grating for the reading beam, whose wavefront after
penetrating the condensate is given by
\begin{equation}
\varphi({\bf r},z_{c+},\tau)=\exp[-i g_{rd}|\phi({\bf r})|^2\tau /\hbar]
\varphi({\bf r},z_{c-},0).
\label{read}
\end{equation}
Here $\tau$ denotes the time it takes the probe atoms to pass through
the condensate of length $l_z$, $z_{c-}$ and $z_{c+}$ are the $z$-coordinates
just before and past the condensate respectively.
The situation described by Eq.\ (\ref{read}) is reminiscent
of phase holography in optics. In case
\begin{equation}
g_{rd} \max[|\phi({\bf r})|^2] \tau/\hbar\ll 1
\label{RNcond}
\end{equation}
we obtain $$\varphi({\bf r},z_{c+},\tau)\simeq (1-
i g_{rd}|\phi({\bf r})|^2\tau/\hbar)\varphi({\bf r},z_{c-},0),$$
i.e., the holographic information stored in the condensate is
indeed transferred to the reading beam. The subsequent free space
propagation allows to separate the different terms contained in
$|\phi({\bf r})|^2$ and to reconstruct the atom-optical replica of the stored
object.

\section{Example: Atom-optical imaging of a small aperture}

In this section we illustrate the principle of atom holography in the case of
imaging of a simple object. This example allows one to investigate
more closely under which conditions and to which degree the general scheme
of Section II can be realized in practice.

\subsection{Optical potentials}

The geometry we are considering is shown in Fig.\ 1b. The aperture and
the condensate are parallel to each other, their centers being located at
the points $(0,0,0)$ and $(0,0,z_c)$, respectively.
The aperture is illuminated by a plane optical wave $E_{inc}$
of amplitude ${\cal E}_0$ and wave vector $k_L$ propagating
along the $z-$direction. The emerging electric field is the well-known
Kirchhoff's solution to the associated diffraction problem \cite{VelRey67}
\begin{eqnarray}
E_o({\bf r},z_c)&=&-\frac{1}{2\pi}\int_{obj}d^2{\bf r}_0\,E_{inc}({\bf r}_0,0)
\left(ik_L-\frac{1}{R}\right)\times\nonumber\\
&& \cos \theta \frac{\exp(ik_LR)}{R},
\label{Kirh}
\end{eqnarray}
where ${\bf r}_0$ is a two-dimensional vector in the object plane and
$R=\sqrt{|{\bf r}-{\bf r}_0|^2+z_c^2}$ the distance from an object
point $({\bf r}_0,z=0)$ to a point $({\bf r},z_c)$ on the thin condensate
acting as a recording medium. Finally, $\cos\theta=z_c/R$, and the
integration is performed over the object boundaries.

The diffracted electric field acquires a simpler form in the Fresnel regime
(i.e., the paraxial or parabolic approximation) where $z_c\gg \lambda_L
=2\pi/k_L$, $|{\bf r}-{\bf r}_0|$ so that $\cos \theta\sim 1$ and
$R\sim z_c+|{\bf r}-{\bf r}_0|^2/z_c$. The object field at the location of
the condensate can then be approximated as
\begin{eqnarray}
E_o({\bf r},z_c)&\propto&e^{ik_Lz_c}e^{i\pi {\bf r}^2/\lambda_L z_c}
\nonumber\\
&&{\cal F}_2[{\cal E}_o O({\bf r}_0)e^{i\pi{\bf r}_0^2/\lambda z_c}
]\left |_{{\bbox \rho}={\bf r}/\lambda_L z_c},\right .
\label{fresn}
\end{eqnarray}
where $O({\bf r}_0)$ defines the shape of the object and
${\cal F}_2[]_{\bbox \rho}$ denotes the two-dimensional Fourier
transform with respect to ${\bbox \rho}$.

In order to construct the optical potential $V({\bf r})$ that imprints the
information onto the condensate, the object field $E_o({\bf r},z_c)$ is
interfered as in conventional holography with a plane wave reference
beam of amplitude ${\cal E}_r$ and wave vector ${\bf k}_r=k_L(\sin\beta,
0,\cos\beta)=(k_\perp,0,k_z)$, see Fig.\ 1b. The intensity of the
superposition at the location of the condensate is thus
\begin{eqnarray}
I({\bf r},z_c)&=& |E_o({\bf r},z_c)|^2+|{\cal E}_r|^2
\nonumber \\
&+&[{\cal F}_2[\dots] {\cal E}_r^\star e^{i(k_L-k_z)z_c
-i  k_\perp x }e^{i\pi {\bf r}^2/\lambda_L z_c} +c.c.]
\label{int}
\end{eqnarray}

To make things clear, let us backtrack for a moment and imagine that
instead of a matter-wave hologram, we create an optical hologram from the
intensity distribution (\ref{int}). When illuminating that hologram with the
reading beam $E_{rd}={\cal E}_{rd}\exp{(ik_z z-i k_\perp  x) }$ ,
we see the emergence of three wavefronts: the background wave
$E_{rd}(|E_o|^2+|E_r|^2)$ travelling along the direction $(-k_{\perp},0,k_z)$;
the object wave $E_{rd}E_o E_r^\star$; and the conjugate beam
$E_{rd}E_o^\star E_r$ which constitutes a converging wave front travelling
in the $z$-direction. Upon propagating a distance $z_c$ after the plane of
the hologram, the quadratic phase in the conjugate beam is undone and a
real image is created. Indeed, by applying again Kirchhoff's solution in
the Fresnel approximation of Eq. (\ref{fresn}) one obtains
\begin{eqnarray}
E_{im}({\bf r},2z_c)&\rightarrow& E_{rd}E_0^\star E_r
\propto e^{ik_Lz_c}e^{i\pi {\bf r}^2/\lambda_L z_c}
\nonumber\\
&\times&{\cal F}_2[E_{rd}E_0^\star E_r
e^{i\pi{\bf r}_0^2/\lambda z_c}
]\left |_{{\bbox \rho}={\bf r}/\lambda_L z_c},\right .
\label{image}
\end{eqnarray}
which is precisely proportional to $O({\bf r})$. The object wavefront,
on the other hand, corresponds to a virtual image. We now study the
conditions under which this same procedure can be applied to atom holography.

\subsection{The writing process}

We assume for concreteness that the condensate consists of sodium atoms, so
that laser fields with wavelengths of about $\lambda_L\sim 10^{-6}$m can
be used to create the optical potentials \cite{StaAndChi98}. It is assumed
to be trapped in a square well potential, as this provides a homogeneous
density of a condensate and thus avoids distortions of the holographic image.
However, one could also work with the approximately constant density
distribution near the center of a harmonic trap that is very wide in the
transverse directions.

To be specific, we investigate the imaging of a
rectangular aperture of width $D=10 \lambda_L$ located at a distance
$z\sim 1000 \lambda_L$ from the condensate. The emerging diffraction
pattern has an angular width of $\theta_d=\lambda_L/D\sim 0.1$, so that
the condensate must have an extension of at least $l_x\sim
100\lambda_L\sim 10^{-4}$m in the $x$-direction. Its extension $l_y$ along
the $y$-axis, as well as the width of the aperture, are
both assumed large enough that diffraction effects are negligible in that
direction, reducing the problem to an effective two-dimensional geometry.
This allows us to express the condensate wave function as
$\phi({\bf r})=\psi(x) /\sqrt{l_yl_z}$ where $l_z$ is the condensate
thickness, which is assumed to be very small as we recall from sec. II C.
Indeed, an upper limit to $l_z$ is provided by the condition that the
density distribution has to be effectively two-dimensional.
The periodicity of the intensity distribution (\ref{int})
along the $z$-direction is determined by the angle between reference and
writing beams; quantitatively, one obtains the requirement
$$
l_z\ll 2\pi/k_L(1-\cos\beta).
$$
In our numerical example we choose $\beta= 30^{\circ}$ and $l_z=10^{-6}$m
so that this condition is well satisfied. From the normalization condition
for a condensate in a square well potential one immediately finds that the
chemical potential is given by
\begin{equation}
\mu=\frac{Ng}{l_xl_yl_z}
\label{mu}
\end{equation}
and thus from Eq.\ (\ref{TFGS})
\begin{equation}
|\psi(x)|^2=\frac{N}{l_x}
\label{dens}
\end{equation}

The strength of the optical fields is determined from the requirement that
the recorded density profile of the condensate is mainly determined by the
light field intensity, and not by the trap ground state profile (pedestal).
This means that the field intensity must yield modulations of the optical
potential deeper than the trapping potential, i.e.
\begin{equation}
\max|\hbar\Omega^2({\bf r},z_c)/\delta|\gg  \mu ,
\end{equation}
where the Rabi frequency $\Omega^2({\bf r},z_c)\propto I({\bf r},z_c)$.
In addition, since the atomic density of Eq. (\ref{TFGS}),
\begin{equation}
|\psi({\bf r},z_c)|^2=\frac{1}{g}{\left[\mu-V_{trap}({\bf r})-
\frac{\hbar\Omega^2({\bf r},z_c)}{\delta}\right]}
\end{equation}
where $V_{trap}$ is the trap potential, is non-negative, one should choose
a negative detuning $\delta$ so that $|\psi({\bf r},z_c)|^2$ reaches the
approximate value
\begin{equation}
|\psi({\bf r},z_c)|^2 \simeq \left|\frac{\hbar\Omega^2({\bf r},z_c)}
{g\delta}\right|.
\label{TFGS1}
\end{equation}
For the Thomas-Fermi approximation and thus Eq.\ (\ref{TFGS}) to be valid,
the condensate healing length $\xi$ needs to be much smaller than the
characteristic length scale $\lambda$ over which the external potential
varies, i.e.,
$$
\xi=(8\pi a n)^{-1/2} \ll \lambda
$$
where $n$ is the condensate density. In our case, $\lambda$ is of course
of the order of the optical writing beam wavelength, $\lambda \simeq
\lambda_L$. The healing length determines the length scale of a density
variation whose quantum pressure (kinetic energy contribution) is of the
order of the interaction energy \cite{DalGioPit98}. This leads to the
requirement
\begin{equation}
8\pi N a l_x/l_y l_z \gg (l_x/\lambda_c)^2
\end{equation}
which, with all other values previously fixed, translates into $N/l_y
\gg 2\times 10^{9}$. In our numerical example we work with the value
$N/l_y=2\times 10^{11}$ which correspond to $N=2\times
10^7$ for $l_y=10^{-4}$m.

We proceed by first determining numerically the ground state of the
condensate subject to the writing optical potentials, with the goal of
justifying the approximate density profile  (\ref{TFGS1}). This is achieved
by solving the Hartree-Fock wave function evolution governed by the
Gross-Pitaevskii equation in imaginary time. For the parameter values listed,
we find a very good agreement between the shape of the density modulation
of the condensate ground state, see Fig.\ \ref{fig2}, and the optical
intensity distribution, see Fig.\ \ref{fig3}.
Further numerical simulations show that this density profile
can actually be prepared by adiabatically turning on the writing and reference
beams: Starting from the condensate in the trap ground state, the optical
fields are switched on slowly enough that no density oscillations become
significantly excited. In our specific example, we consider first the
square-well ground state $\psi(x)=\sqrt{N/l_x}$
and turn on the optical potential Eq.\ (\ref{int}) with a switch-on function
$[1-\exp(-t/t_c)]^2$, where $t_c\simeq 0.01$s is short in comparison to
typical condensate lifetimes. The resulting condensate density profile is
given in Fig.\ \ref{fig4}. This illustrates that in this way the condensate
wave function is transferred without difficulty from the trap ground state
to the new ground state in presence of the optical
potential, see Fig.\ \ref{fig3}.

\centerline{\psfig{figure=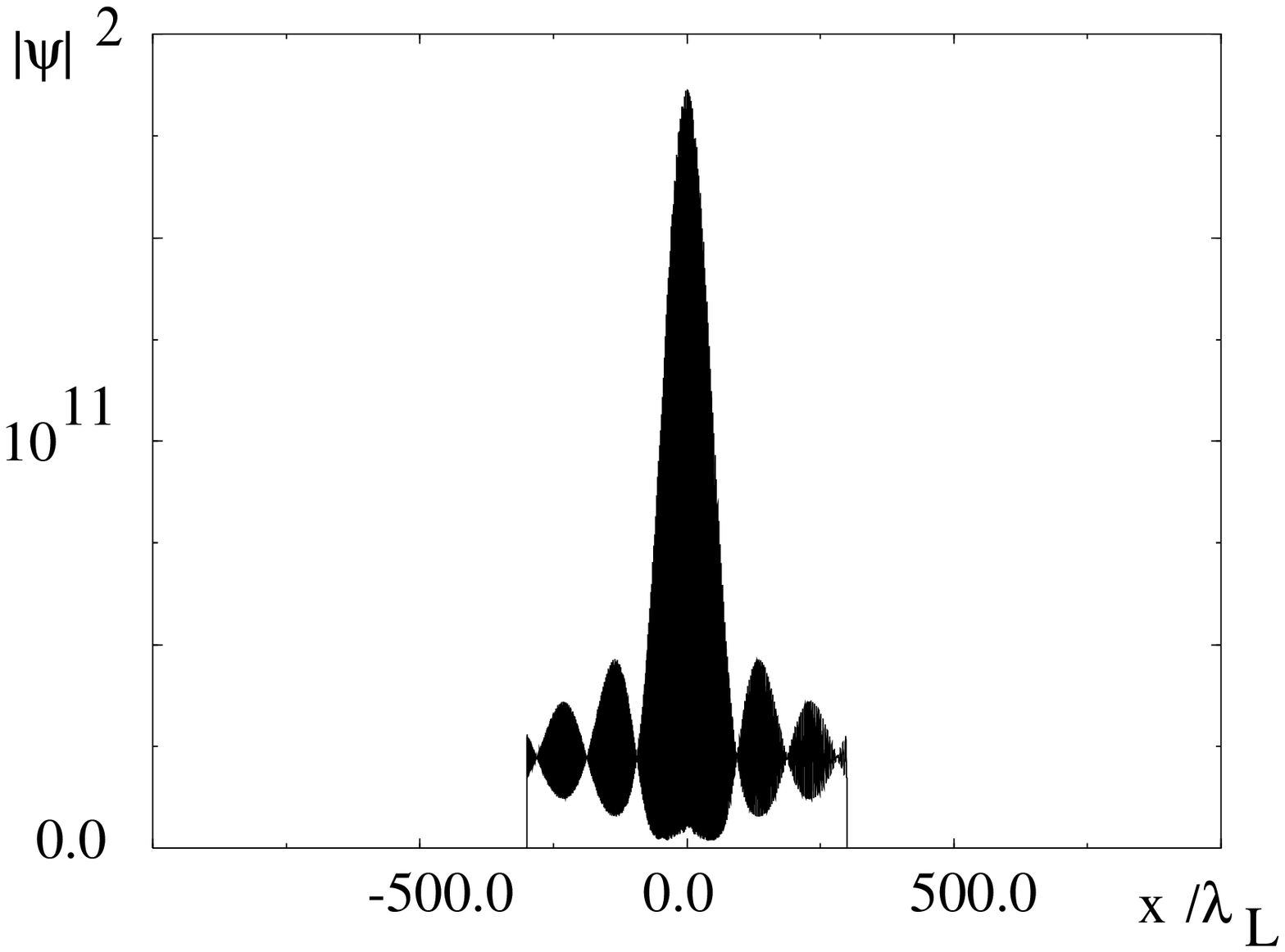,width=8.6cm,clip=}}
\begin{figure}
\caption{Condensate ground state density [m$^{-1}$] in a square-well potential
with the optical fields on. The object size is $ 10\lambda_L$, and the distance
from the object to the condensate is $z_c=1000\cdot \lambda_L$. The sodium
condensate parameters are $N=2\cdot 10^{7}$, $l_{x}=3\cdot 10^{-4}$m,
$l_y=10^{-4}$m,  $l_z=10^{-6}$m, $V_{trap}(|x|>l_{x})=\mu$. The
optical field parameters are $|\hbar\Omega^2({\bf r},z_c)/\delta|
=100\cdot \mu$, ${\cal E}_r=0.1{\cal E}_o$, $\beta=30^0$}
\label{fig2}
\end{figure}

\centerline{\psfig{figure=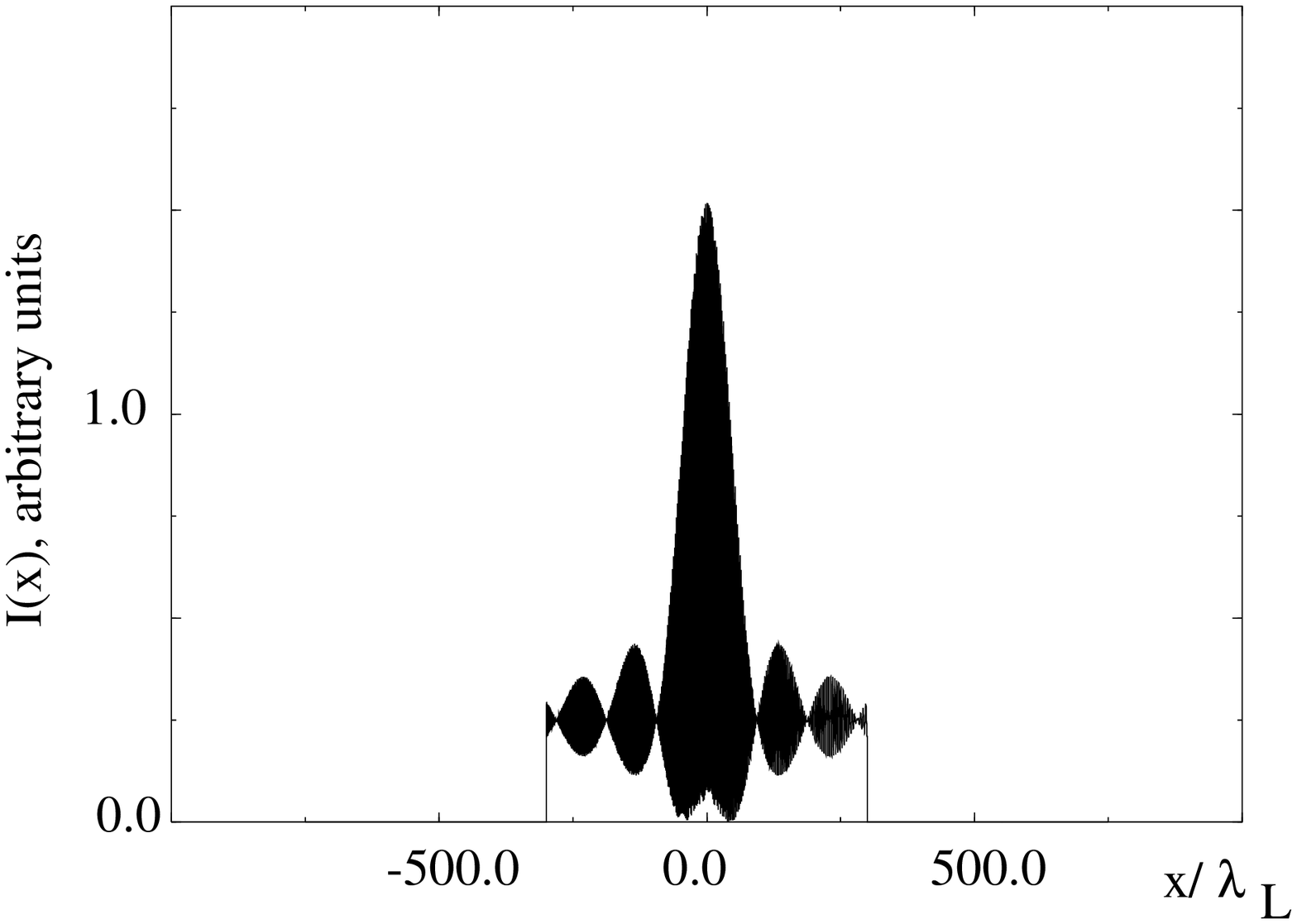,width=8.6cm,clip=}}
\begin{figure}
\caption{Optical hologram for the same optical fields as used for
 writing on the atomic condensate in Fig.\ \ref{fig2}.}
\label{fig3}
\end{figure}

\subsection{The reading process}

The reading and construction of a replica of the store object is achieved by
an off-axis atomic beam impinging upon the condensate from the side opposite
to the writing optical beams. The velocity of this beam needs to be
carefully selected, as it must be confined between a lower and an upper
bound resulting from the thin hologram and $s$-wave scattering approximations,
respectively.

As we have seen, a lower bound in atomic
velocities $v_{rd}$ in that beam is determined by the condition of validity of
Raman-Nath, or thin hologram, approximation. The physical meaning of this
condition is that the longitudinal
\centerline{\psfig{figure=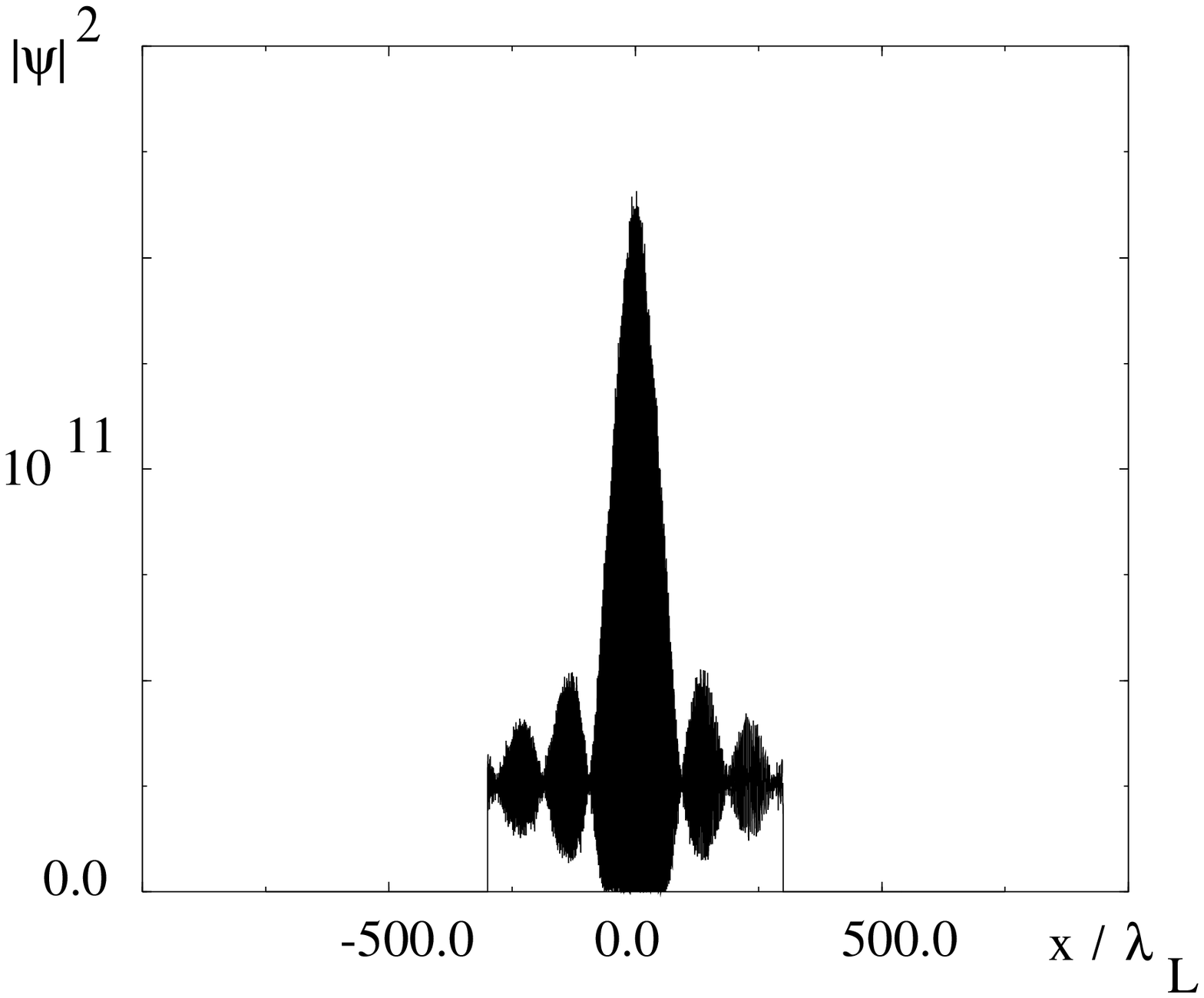,width=8.6cm,clip=}}
\begin{figure}
\caption{Steady-state atomic density  profile [m$^{-1}$] for the system of
 Fig.\ \ref{fig2}, but achieved by adiabatically turning on the optical
 potential in a characteristic time $t_c=0.01 sec$}
\label{fig4}
\end{figure}
\noindent
contribution to the kinetic energy of the probe atoms must be large
compared to their interaction energy with the condensate atoms
\begin{equation}
\frac{p_z^2}{2 M_{rd}}\gg g_{rd}max|\phi({\bf r})|^2
\label{ineq2}
\end{equation}
and the typical transverse deflection inside the condensate must remain
small compared to the length scale of the density fluctuations,
\begin{equation}
\frac{p_x \tau}{M_{rd}}\ll \lambda_L.
\label{ineq3}
\end{equation}
Under these circumstances, the kinetic energy term in
Eq.\ (\ref{Schr}) can be dropped. Eq.\ (\ref{ineq2}), together with the
requirement of small phase variations in the reading beam upon propagation
through the condensate (see Eq.\ (\ref{RNcond})), gives the lower bound for
$v_{rd}$.

In addition, Eq.\ (\ref{ineq3}) together with the requirement that
the $s$-wave scattering approximation is valid, see Eq. (\ref{scat_cond}),
determines an upper bound for $v_{rd}$.

Our numerical simulations are for $v_{rd}\sim 10^{-1}$ m/sec, which is a
reasonable value for the experiments with ultra cold atomic beams and lies
within these lower and upper bounds.

Similarly to the optical case, the atomic wave function acquires a quadratic
phase upon free propagation, a result of the fact that in the paraxial
approximation, the dispersion relations of optical and matter waves are
both quadratic and essentially the same. Indeed,
\begin{eqnarray}
& &\varphi(x,z_{c+}+\Delta z,\tau+\Delta t)=
\nonumber\\ & &
\int d\xi  e^{i x \xi} e^{ -i\frac {\hbar \xi^2} {2 M_{rd}} \Delta t }
\int dx'  e^{-i x \xi} \varphi(x',z_{c+},\tau)  \nonumber \\
&&= e^{ i\frac {M_{rd}} {2 \hbar \Delta t} x^2 }
\int dx' \varphi(x',z_{c+},\tau)
e^ {i\frac{M_{rd}}{2 \hbar \Delta t}x'^2}
e^{-i\frac{M_{rd}}{\hbar \Delta t}x x'} \nonumber \\
&&=e^{i\frac{M_{rd}}{2 \hbar \Delta t}x^2} {\cal F}_2 [\varphi(x',z_{c+},\tau)
e^{i\frac{M_{rd}}{2 \hbar \Delta t}x'^2}]\left
|_{\xi=\frac{M_{rd}}{\hbar \Delta t}x}\right .
\label{atom_rd}
\end{eqnarray}
Here $\phi(x',z_{c+},\tau)$ is defined according with Eq.\ (\ref{read})
and $\Delta z = v_{rd}\Delta t$. The free propagation time $\Delta t$ is
determined from the same requirement as in the optical case, namely that
the quadratic phase $\exp[-i\pi x^2/\lambda_L z_c]$ in the recorded atomic
density (\ref{TFGS1}) be exactly compensated. This gives
\begin{equation}
\Delta t = \frac{M_{rd}}{2\hbar} \left (\frac{\lambda_L z_c}{\pi} \right ) .
\end{equation}
We finally observe that in order for the image to be formed on-axis,
the reading beam needs to propagate off-axis at an angle $\beta_A$ such
that it compensates the angle of an optical reference beam, i. e.
\begin{equation}
\sin\beta_A=k_L\sin\beta/k_A .
\end{equation}
This is a very small angle since $k_A\equiv v_{rd}M_{rd}/\hbar \gg k_L$.
Fig.\ \ref{fig5} shows the numerically computed atomic density profile in the
plane $z_c+\Delta z$ for this choice of parameters. This demonstrates
explicitly that the original rectangular aperture is indeed reconstructed
on axis. In contrast, the virtual image, for which the quadratic phase
stored in the condensate is not compensated, propagates off-axis, and so does
the background contribution. \\

\centerline{\psfig{figure=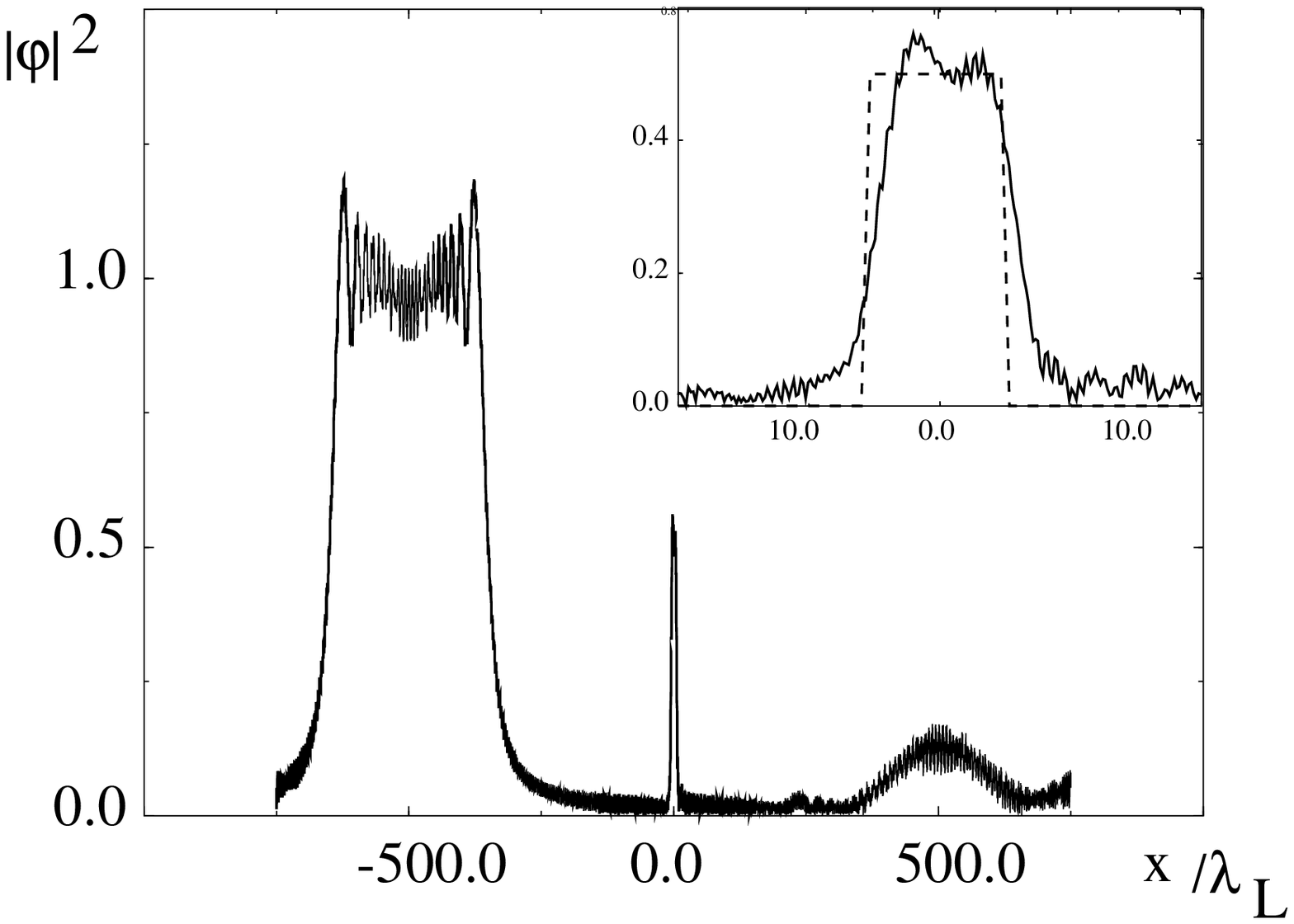,width=8.6cm,clip=}}
\begin{figure}
\caption{Reconstructing a replica of the original object from the atomic
 hologram. The reading beam consists of a monochromatic beam of sodium atoms
 moving at an angle $\beta_A$ from the $z$-axis at a velocity
 $v_{rd}=0.1$ m/sec. Shown is the atomic density profile at a distance
 $\Delta z$ from the condensate such that the quadratic phase shift of the
 conjugate image is precisely canceled. The insert compares the reconstructed
 and original objects. The off-axis feature for positive $x$ corresponds to
 the real object, for which the quadratic phase is still present. The
 large off-axis feature at negative $x$ is background.}
\label{fig5}
\end{figure}
\noindent

\section{Summary and Conclusions}
In this paper, we have theoretically discussed an atom holography scheme
where the hologram is stored in a Bose-Einstein condensate. An
important feature of the proposed scheme is that the reading beam needs not
consist of the same element as the condensate atoms. It merely needs to be a slow
monochromatic atomic or molecular beam that interacts with the condensate
atoms via $s$-wave scattering, or any other interaction leading to a cubic
nonlinearity in the nonlinear Schr\"odinger equation. Matter-wave holography,
once experimentally realized, is certain to open up the way to numerous
potential applications, in particular in microfabrication. One should note
that the slow atoms discussed in the present paper have large de Broglie
wavelengths, which severely limit the spatial resolution of the material
replica that can be reconstructed. However, the wavelength of matter waves
can easily be shortened, for example by gravitation. It is readily
conceivable that this can be used to achieve miniaturized structures with
nanometer-scale features. This, and other aspects of atom holography, will
be the subject of future studies.

\acknowledgements
The authors aknowledge numerious discussions and valuable suggestions from
M. G. Moore. This work is supported in part by the U.S.\ Office of Naval Research
under Contract No.\ 14-91-J1205, by the National Science Foundation under
Grant No.\ PHY98-01099, by the U.S.\ Army Research Office, and by the
Joint Services Optics Program.

\end{document}